# Efficient Hybrid White Organic Light-Emitting Diodes for Application of Triplet Harvesting with Simple Structure

Kyo Min Hwang, Song Eun Lee, Sungkyu Lee, Han Kyu Yoo, Hyun Jung Baek

and Young Kwan Kim*

*Department of Information Display, Hongik University, Seoul 121-791*

Jwajin Kim and Seung Soo Yoon**

*Department of Chemistry, Sungkyunkwan University, Suwon 440-746*

In this study, we fabricated hybrid white organic light-emitting diodes (WOLEDs) based on triplet harvesting with simple structure. All the hole transporting material and host in emitting layer (EML) of devices were utilized with same material by using N,N'-di-1-naphthalenyl-N,N'-diphenyl-[1,1':4',1'':4'',1'''-quaterphenyl]-4,4'''-diamine (4P-NPD) which were known to be blue fluorescent material. Simple hybrid WOLEDs were fabricated three color with blue fluorescent and green, red phosphorescent materials. We was investigated the effect of triplet harvesting (TH) by exciton generation zone on simple hybrid WOLEDs. Characteristic of simple hybrid WOLEDs were dominant hole mobility, therefore exciton generation zone was expected in EML. Additionally, we was optimization thickness of hole transporting layer and electron transporting layer was fabricated a simple hybrid WOLEDs. Simple hybrid WOLED exhibits maximum luminous efficiency of 29.3 cd/A and maximum external quantum efficiency of 11.2%. Commission Internationale de l'Éclairage (International Commission on Illumination) coordinates of (0.45, 0.43) at about 10,000 cd/m$^2$.

\*Email: kimyk@hongik.ac.kr

\*Fax: +82-2-320-1646

\*\*Email: ssyoon@skku.edu

\*\*Fax: +82-31-290-7075

# I. INTRODUCTION

White organic light-emitting diodes (WOLEDs) have obtained significant attention in recent years due to their potential applications in high-resolution, full-color displays, wide view angle, flexibility and solid-state lighting [1-3]. State of the art phosphorescent WOLEDs show outstanding efficiencies as all electrically generated singlet and triplet excitons can be harvested with phosphorescent material [4,5]. However, the lack of blue phosphorescent material with high operating stability limits the further development of phosphorescent WOLEDs [6,7]. This limitation can be improved through the use of high stability blue fluorescent material and high efficiency green, red phosphorescent, rendering these hybrid structures better commercialization anticipates for WOLEDs [8,9]. For lighting manufactured goods, the use of two colors to realize white light suffers a low grade color-rendering index (CRI), which was unsatisfactory for white light sources, making this approach less charming for generating high value white light sources. Therefore, it was essential to constitute a three colors structure to obtain a spectrum of the visible region in hybrid WOLEDs.

To achieve high efficiency for hybrid WOLEDs, mutual quenching of fluorescent and phosphorescent emitter should be prevented in emitting layer (EML) [10]. To overcome, device structure consists of an interlayer between the fluorescent and the phosphorescent layers reduced the triplet exciton loss [11,12]. However, the use of the interlayer will restrict the device efficiency [10]. Firstly, the voltage drop across the interlayer is nothing to sneeze, leading to the lower power efficiency. Secondly, the addition of an layer brings additional interfaces which increase the possibility of exiplex formation will impair the quantum efficiency of hybrid WOLEDs [13]. Lastly, with the increase of interlayer in EML, the driving voltage of an OLED is increased the devices [14]. To date, there have been no reports on three color hybrid WOLEDs with simple structure. Also, the triplet harvesting (TH) contributed to light emission for triplet excitons to non-radiative decay in fluorescent EML, which can lead white emission to a wide spectral range and to enhance efficiency of the hybrid WOLEDs.

In this paper, we propose a structure for hybrid WOLEDs that does not follow the conventional structure with an interlayer between blue fluorescent and red, green phosphorescent layer. Our concept was to introduce simple structure with p-type blue fluorescent material as a N,N'-di-1-naphthalenyl-N,N'-diphenyl-[1,1':4',1'',4'',1'''-quaterphenyl]-4,4'''-diamine (4P-NPD) to alter the hole transport layer (HTL). Moreover, non-radiative decay of triplet excitons utilize TH mechanism in simple hybrid WOLEDs. For this purpose, we fabricated for reducing the number of layers, which applied to stable and simple hybrid WOLEDs structure.

## II. EXPERIMENTS

To fabricate a simple hybrid WOLEDs, indium-tin-oxide (ITO) coated glass was prepared in an ultrasonic bath using the acetone, methyl alcohol, deionized water, and ethyl alcohol, respectively. ITO has sheet resistance of ~10 $\Omega$/sq and the emitting area was $3 \times 3$ mm$^2$. Pre-cleaned ITO was treated by oxygen plasma for 2 min under a low vacuum of $2 \times 10^{-2}$ torr. All organic layers were deposited in thermal evaporation under a high vacuum of $5 \times 10^{-7}$ torr. The deposition rate of organic layers was 1 Å/s and lithium quinolate (Liq) was 0.1 Å/s and the deposition rate of aluminum (Al) electrode was 10 Å/s. After the evaporation, to prevent moisture, getter was used to in the encapsulation.

We fabricated several hybrid WOLEDs with various simple structures, including emitter materials with the following chemical structures, a blue fluorescent emitter 4P-NPD, a red phosphorescent emitter iridium(III) bis(2-phenylquinoline) acetylacetonate (Ir(pq)$_2$acac) and a green phosphorescent emitter tris(2-phenylptridine)iridium (Ir(ppy)$_3$) (figure 1(a)). We used tris(4-carbazoyl-9-ylphenyl)amine (TCTA) and 1,3,5-tris(N-phenylbenzimidazole-2-yl)benzene (TPBi) as HTL and electron transporting layer (ETL). The notable point, TCTA of HTL was replaced by 4P-NPD. Effect that can be reduced the interface in simple hybrid WOLEDs. Figure 1(b) shown simple hybrid WOLEDs structure and summarized in Table 1. We controlled the HTL and ETL thickness of organic layers to confirm a capability of charge confinement for realization of highly efficient simple hybrid WOLEDs structure.

Optical properties and electrical properties of the fabricated device (voltage-luminance characteristics (V–L), luminous efficiency (LE), external quantum efficiency (EQE), the electroluminescence spectra, Commission International de L'Éclairage (CIE) chromaticity coordinate) were measured by using Keithley 2400 and Chroma meter CS-1000A instruments, respectively.

## III. RESULTS AND DISCUSSION

We were schematized operated carrier within device and transferred triplet exciton within exciton generation zone, as shown in the figure 2. Hole mobility of TCTA and 4P-NPD were $\mu_h$(TCTA) : 3.0 × $10^{-4}$cm$^2$/V s and $\mu_h$(4P-NPD) : 6.6 × $10^{-4}$cm$^2$/V s [15,16]. For reason of similar hole mobility, we was substitute 4P-NPD with TCTA for less material and can be produced a simple hybrid WOLEDs structure. Moreover, the predominant hole injection by exciton generation zone was assumed to be located to the 4P-NPD of 3 nm. According to spin statistics, 25% singlet excitons and 75% triplet excitons were generated. The 25% of singlet excitons were used for the fluorescent emission, whereas 75% of triplet excitons were non-radiative decay in fluorescent EML. Using the TH mechanism, it was possible to phosphorescent emission for blue fluorescent triplet excitons energy transfer to triplet energy state of adjacent green and red phosphorescence. Effect of TH concept can lead white emission to a wide spectral range and to enhanced efficiency of the hybrid WOLEDs. Non-radiative triplet excitons of 4P-NPD (triplet energy level ($T_1$) : 2.3 eV) will facilitate an efficient TH mechanism to green ($T_1$ : 2.4 eV) and red ($T_1$ : 2.2 eV) phosphorescent materials [17-19].

Figure 3(a) and inset shows the luminous efficiency–luminance (LE–L) and external quantum efficiency–luminance (EQE–L) of fabricated simple hybrid WOLEDs (device A, B, C, and D). The relatively low turn-on voltage of 2.53 V, 2.49 V, 2.47 V, and 2.50 V were observed for the device A, B,

C, and D, respectively. The reduced thickness of the 4P-NPD turn-on voltage was faster, while turn-on voltage of device D was increased due to the imbalance of the carrier injection [20]. At a given constant luminance of 10,000 cd/m$^2$, the LE and EQE were 15.4 cd/A and 6.8% for the device A, 17.2 cd/A and 7.5 % for the device B, 17.8 cd/A and 7.8% for the device C, and 16.4 cd/A and 6.9% for the device D. Figure 3(b) shows the normalized electroluminescence (EL) intensity of simple hybrid WOLEDs (device A, B, C, and D). The EL spectra of simple hybrid WOLEDs exhibit warm white light emission comprising distinct blue fluorescent emission and green, red phosphorescent emission covering the range from 400 to 700 nm. As shown in the inset of figure 3(b), the CIE coordinates shown an unvaried from (0.48, 0.41) at 1,000 cd/m$^2$ to (0.46, 0.40) at 10,000 cd/m$^2$. Based on these results, it was set to the optimized thickness of the HTL as 20 nm.

We need try to ensure that generated exciton in 4P-NPD can be apply TH mechanism to adjacent green, red phosphorescent. Figure 4(a) shown device C and the exciton blocking layer (EBL) devices structure. EBL was selected for material that can be triplet exciton blocking between phosphorescent and fluorescent EML and not influence as much as possible in the structure. It was used TCTA that selected same highest occupied molecular orbital (HOMO), lowest unoccupied molecular orbital (LUMO) of 4P-NPD and high $T_1$ (2.8 eV). TPBi was selected EBL near the ETL side of EML, because that was high $T_1$ (2.6 eV) and used as an ETL (figure 2) [21,22]. Due to the insertion of the EBL, TH was not performed on direction in device from 4P-NPD to Ir(pq)$_2$acac or Ir(ppy)$_3$. Emissions of excitons were shown the area to be expected. Figure 4(b) was shown the EL spectra. Green intensity was obtained strong graph in EBL device A better than device C, red intensity was obtained strong graph in EBL device B better than device C. Therefore, triplet excitons of fluorescence were ensured TH to the adjacent layer and it was able to prevent TH by EBL.

The thickness of ETL was optimized to create a simpler structure. Device C was ETL as 30 nm that produced previously, device E and F were ETLs as 40 nm and 20 nm. Figure 5(a) and inset shown the power efficiency–L (PE–L) and EQE–L of fabricated device C, E, and F. Performance of device F was the lowest efficiency, because to injection of electron becomes faster compared to the other device by thin ETL thickness. There appears to be an imbalance of carrier injection. In contrast, ETL as 40 nm was injection of electron becomes slower than device C and F. Accordingly, there appears to be an imbalance of carrier injection due to electron injection. As shown in the figure 5(b) can be demonstrated to the best color distribution of device C appears better than any other devices. If the injection of electron was slow, it was possible to see the dominant green intensity and in contrast that was fast, it was possible to see the dominant red intensity.

## IV. CONCLUSION

In summary, we report the simple hybrid WOLEDs structure with TH mechanism. The blue fluorescent material as 4P-NPD showed hole dominant charge transporting behavior. Accordingly, it could be replaced TCTA as 4P-NPD, therefore we reduced interface in simple hybrid WOLED. In addition, simple hybrid WOLEDs was compared with the devices for controlling the thickness of HTL and ETL. Device C showed best performances, because of improved carrier balance and exciton distribution within EML. The simple hybrid WOLEDs was fabricated that utilized simple structure and TH. Optimization simple hybrid WOLED properties of low driving voltage of 2.46 V, maximum EQE up to 11.2%, and CIE coordinated of (0.45, 0.43). Obviously and such notified output would apply a useful guide for high-performance simple hybrid WOLEDs, which would be advantageous for the realization of full color flat-panel displays and lighting in the future.

## ACKNOWLEDGEMENT

This research was supported by Basic Science Research Program through the National Research Foundation of Korea (NRF) funded by the Ministry of Education (No. 2015R1A6A1A03031833) and the MSIP(Ministy of Science, ICT and Future Planning), Korea, under the ITRC(Information

Technology Research Center) support program (IITC-2016-H8501-16-1009) supervised by the IITP(Institute for Information & communications Technology Promotion).

**REFERENCES**

[1] T. C. Rosenow, M. Furno, S. Reineke, S. Olthof, B. Lussem and K. Leo, *J. Appl. Phys.* **108**, 113113 (2010).

[2] S. W. Liu, X. W. Sun and H. V. Demir, *Aip. Adv.* **2**, 012192 (2012).

[3] S. Y. Lee, T. Yasuda, H. Nomura and C. Adachi, *Appl. Phys. Lett.* **101**, 093306 (2012).

[4] Z. Y. Xie and L. S. Hung, *Appl. Phys. Lett.* **84**, 1207 (2004).

[5] S. Reineke, F. Lindner, G. Schwartz, N. Seidler, K. Walzer, B. Lussem and K. Leo, *Nature* **459**, 234 (2009).

[6] Y. Duan, M. Mazzeo, V. Mariano, F. Mariano, D. Qin, R. Cingolani and G. Gigli, *Appl. Phys. Lett.* **92**, 113304 (2008).

[7] M.C. Gather, A. Kohnen and K. Meerholz, *Adv. Mater*. **23**, 233 (2011).

[8] H. Sasabe, K. Minamoto, Y. J. Pu, M. Hirasawa and J. Kido, *Org. Electron.* **13**, 2615 (2012).

[9] R. Seifert, I. R. D. Moraes, S. Scholz, M. C. Gather, B. Lussem and K. Leo, *Org. Electron.* **14**, 115 (2013).

[10] G. Schwartz, S. Reineke, T. C. Rosenow, K. Walzer and K. Leo, *Adv. Funct. Mater.* **19**, 1319 (2009).

[11] M. E. Kondakova, J. C. Deaton, T. D. Pawlik, D. F. Giesen, D. Y. Kondakov, R. H. Young, T. L. Royster, D. L. Comfort and J. D. Shore, *J. Appl. Phys.* **107**, 014515 (2010).

[12] D. Zhang, L. Duan, Y. Li, D. Zhang and Y. Qiu, *J. Mater. Chem. C* **2**, 8191 (2014).

[13] F. Zhao, Z. Zhang, Y. Liu, Y. Dai, J. Chen and D. Ma, *Org. Electron.* **13**, 1049 (2012).

[14] Y. Sun, N. C. Giebink, H. Kanno, B. Ma, M. E. Thompson and S. R. Forrest, *Nature* **440**,

908(2006)

[15] L. Y. Guo, X. L. Zhang, C. Liu, W. Y. Chen, B. X. Mi, J. Song, Y. H. Li and Z. Q. Gao, *Org. Electron.* **15**, 2964 (2014).

[16] J. H. Lee, C. I. Wu, S. W. Liu, C. A. Huang and T. Chang, *Appl. Phys. Lett.* **86**, 103506 (2005).

[17] S. Hofmann, M .Hummert, R. Scholz, R. Luschtinetz, C. Murawski, P. A. Will, S. I. Hintschich, J. Alex, V. Jankus, A. P. Monkman, B. Lussem, K. Leo and M .C. gather, *Chem. Mater.* **26**, 2414 (2014).

[18] C. Diez, T. C. G. Reusch, S. Seidel and W. Brutting, *J. Appl. Phys.* **111**, 113102 (2012).

[19] S. O. Jeon, K. S. Yook, C. W. Joo and J. Y. Lee, *Org. Electron.* **11**, 881 (2010).

[20] T. Zhang, B. Zhao, B. Chu, W. Li, Z. Su, X. Yan, C. Liu, H. Wu, Y. Gao, F. Jin and F. Hou, *Sci. Rep.* **5**, 1 (2015).

[21] J. S. Swensen, E. Polikarpov, A. V. Ruden, L. Wang, L. S. Sapochak and A. B. Padmaperuma, *Adv. Funct. Mater.* **21**, 3250 (2011).

[22] S. C. Lo, R. E. Harding, C. P. Shipley, S. G. Stevenson, P. L. Burn and D. W. Samuel, *J. Am. Chem. Soc.* **46**, 16681 (2009).

**Table 1.** Detailed structures of simple hybrid WOLEDs.

| Device | HTL | EML | ETL | EIL | Cathode |
|---|---|---|---|---|---|
| Device A | TCTA (30 nm) | 4P-NPD:Ir(pq)$_2$acac 0.5% (15 nm) / 4P-NPD (3 nm) / TPBi:Ir(ppy)$_3$ 6% (5 nm) | TPBi (30 nm) | Liq (2 nm) | Al (100 nm) |
| Device B | 4P-NPD (30 nm) | | TPBi (30 nm) | Liq (2 nm) | Al (100 nm) |
| Device C | 4P-NPD (20 nm) | | TPBi (30 nm) | Liq (2 nm) | Al (100 nm) |
| Device D | 4P-NPD (10 nm) | | TPBi (30 nm) | Liq (2 nm) | Al (100 nm) |
| Device E | 4P-NPD (20 nm) | | TPBi (40 nm) | Liq (2 nm) | Al (100 nm) |
| Device F | 4P-NPD (20 nm) | | TPBi (20 nm) | Liq (2 nm) | Al (100 nm) |

**Figure Captions.**

Figure 1. (a) Molecular structure of blue fluorescent material, and green and red phosphorescent material. (b) Energy band diagram and device configuration for simple hybrid WOLEDs (device A, B, C, D, E, and F).

Figure 2. Energy band diagram and exciton energy diagram of TH mechanism in EML.

Figure 3. (a) Luminous efficiency-luminance and (inset) external quantum efficiency-luminance of device A, B, C, and D. (b) Normalized EL spectra and (inset) CIE coordinates of device A, B, C, and D.

Figure 4. (a) Energy band diagram and device configuration for device C, EBL device A, and B. (b) EL spectra of device C, EBL device A, and B measured at 1,000 cd/m$^2$.

Figure 5. (a) Power efficiency-luminance and (inset) external quantum efficiency-luminance of device C, E, and F. (b) Normalized EL spectra of device C, E, and F.

(a)

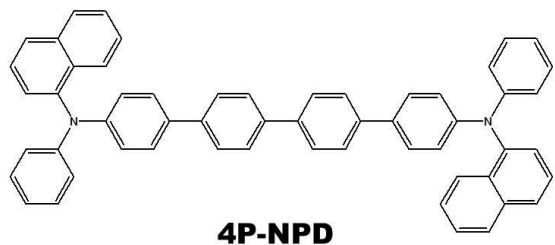

**4P-NPD**

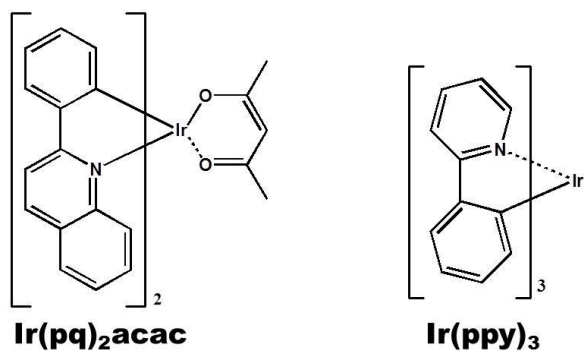

Ir(pq)$_2$acac        Ir(ppy)$_3$

(b)

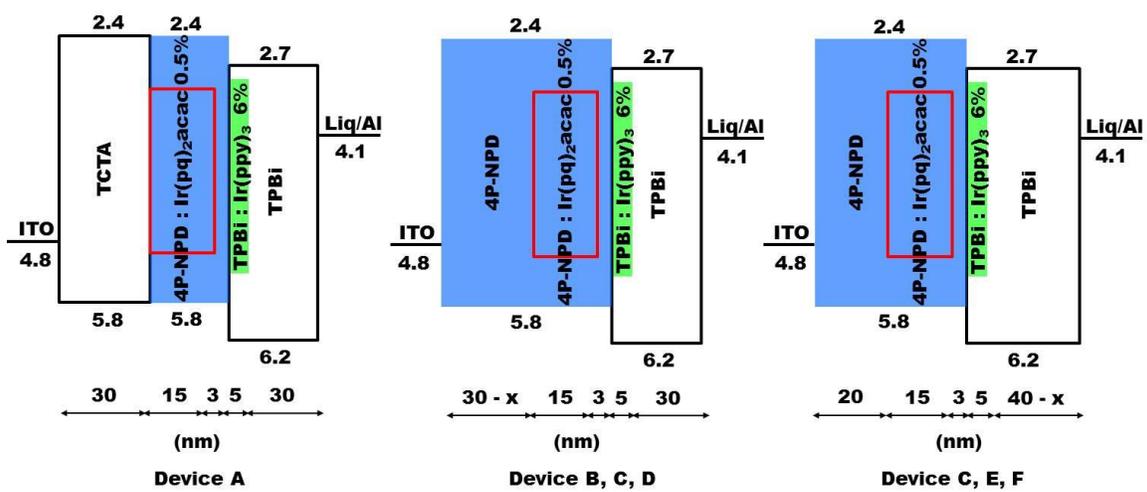

Figure 1

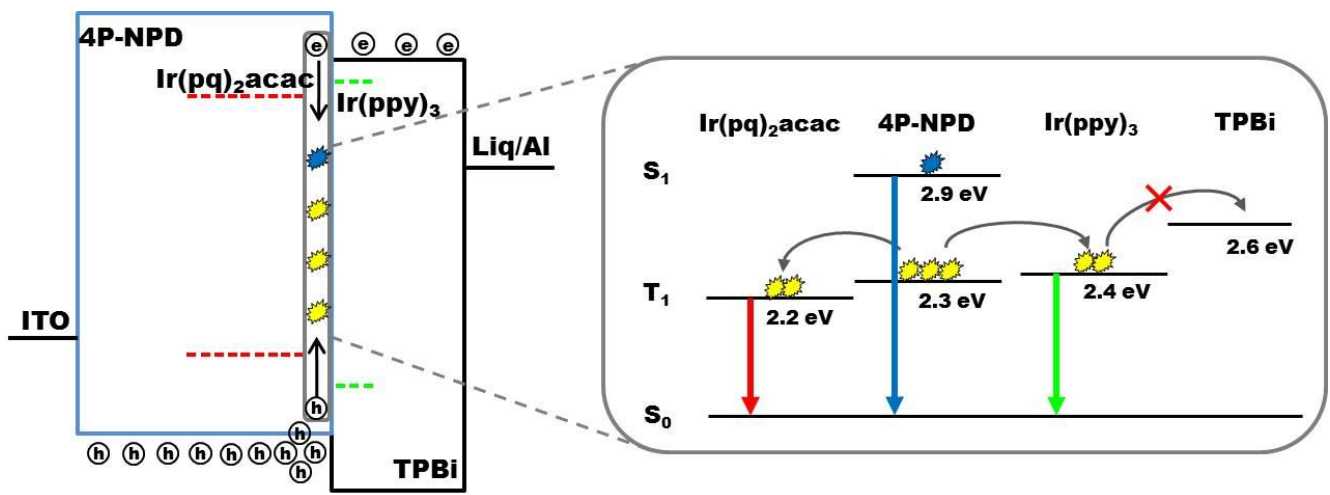

Figure 2

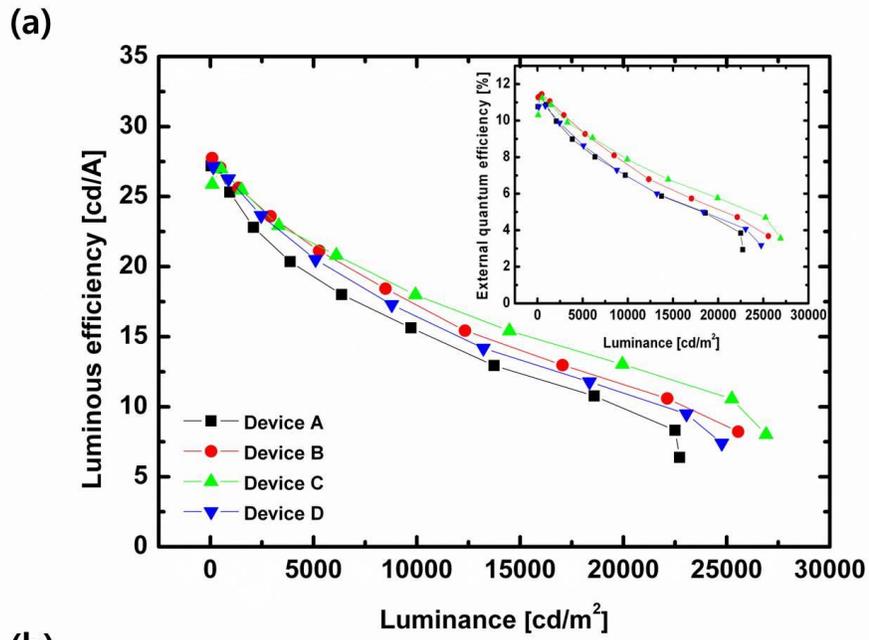
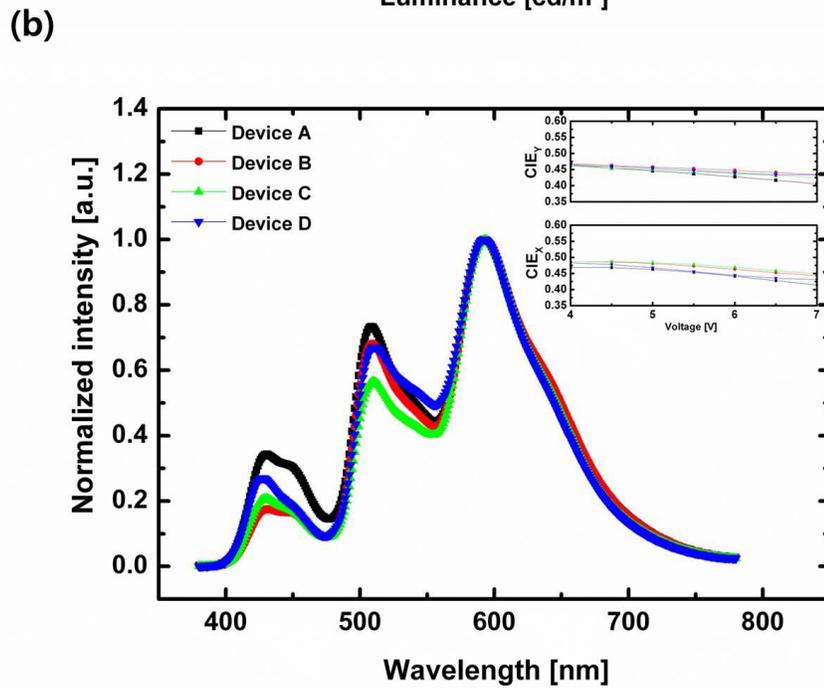

Figure 3

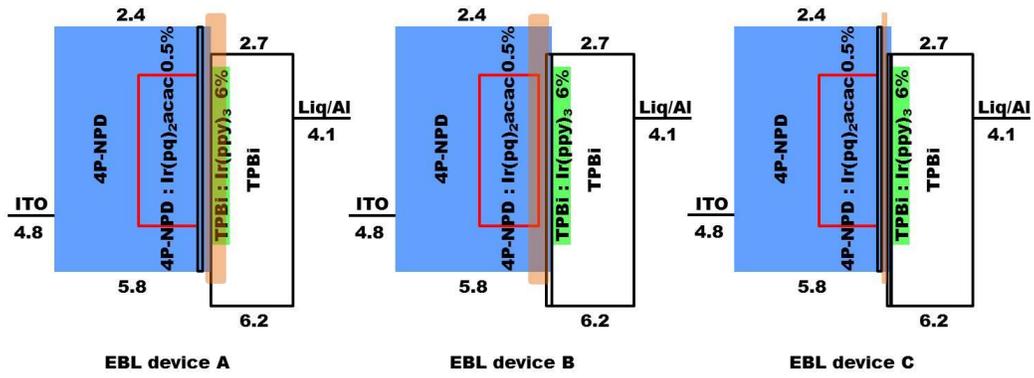

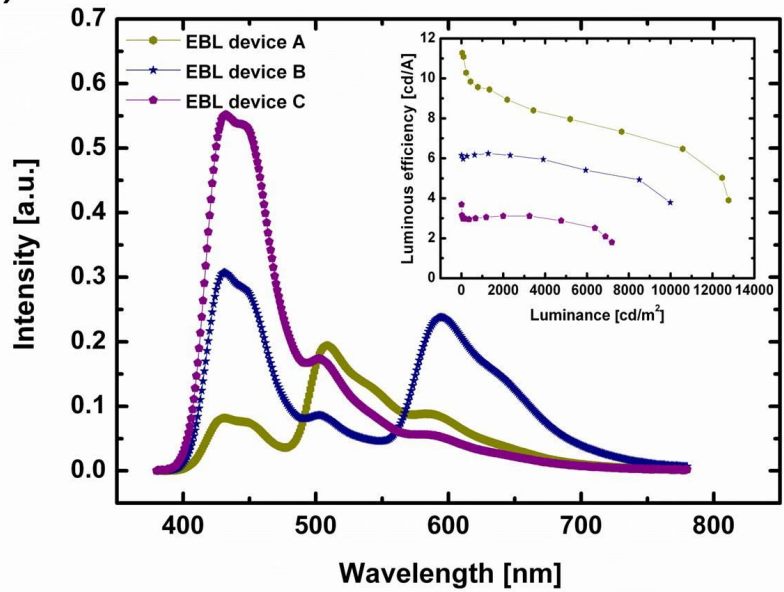

Figure 4

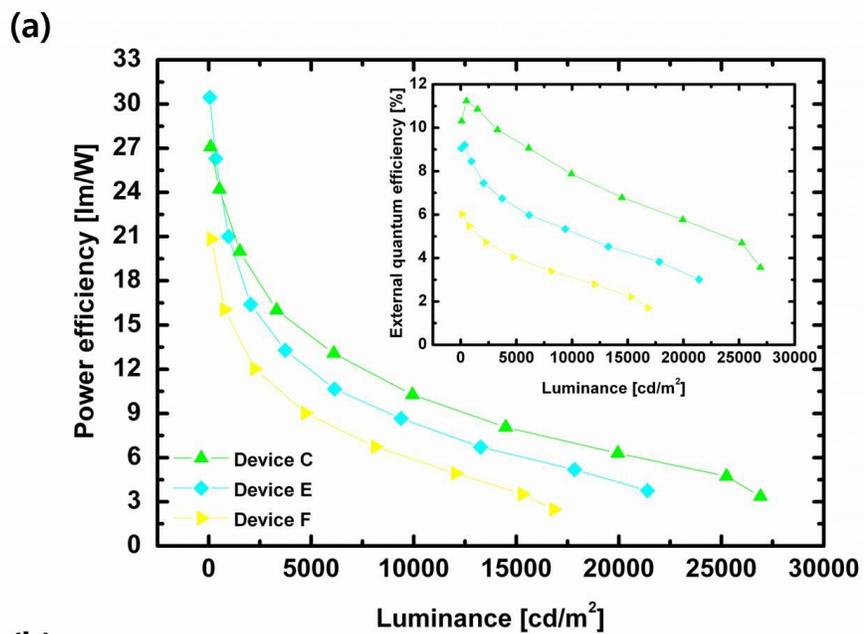

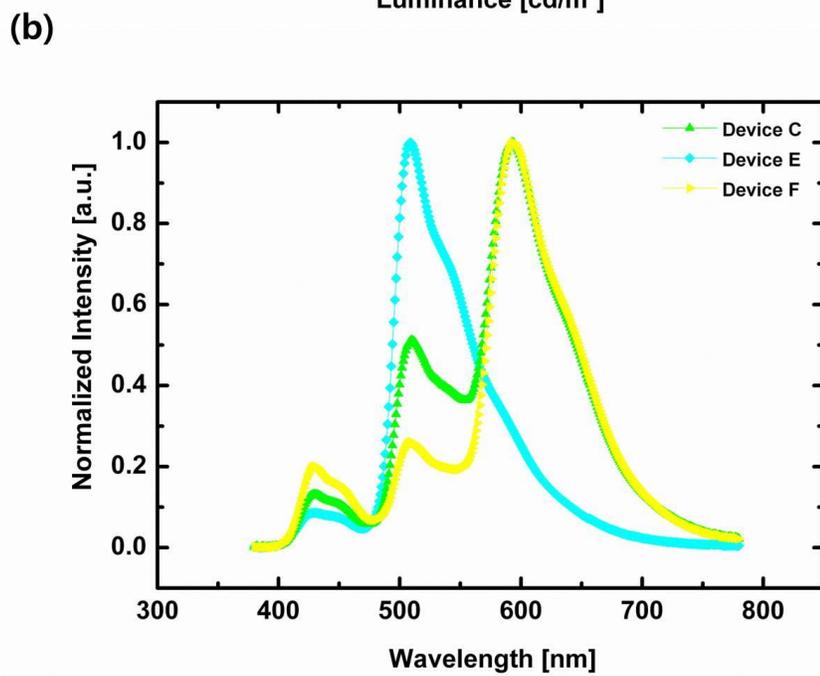

Figure 5